\documentclass[graybox]{svmult}

\usepackage{mathptmx}       
\usepackage{helvet}         
\usepackage{courier}        
\usepackage{type1cm}        
\usepackage{makeidx}         
\usepackage{graphicx}        
\usepackage{multicol}        
\usepackage[bottom]{footmisc}
\makeindex             

\usepackage{url,amsmath,amssymb,placeins}
\usepackage[numbers]{natbib}
\newcommand{\ignore}[1]{}

\begin{document}

\title*{Differential network analysis and\newline graph classification: a glocal approach}
\author{Giuseppe Jurman, Michele Filosi, Samantha Riccadonna, Roberto Visintainer, Cesare Furlanello}
\authorrunning{Jurman \textit{et al.}}
\institute{G. Jurman, M. Filosi, R. Visintainer, C. Furlanello \at Fondazione Bruno Kessler, via Sommarive 18 Povo, I-38122 Trento, Italy\newline\email{{jurman,filosi,visintainer,furlan}@fbk.eu}
\and S. Riccadonna \at Centro Ricerca e Innovazione, Fondazione Edmund Mach, I-38010 San Michele all'Adige, Italy\newline\email{samantha.riccadonna@fmach.it}}
\maketitle

\abstract{
Based on the glocal HIM metric and its induced graph kernel, we propose a novel solution in differential network analysis that integrates network comparison and classification tasks. 
The HIM distance is defined as the one-parameter family of product metrics linearly combining the normalised Hamming distance H and the normalised Ipsen-Mikhailov spectral distance IM. 
The combination of the two components within a single metric allows overcoming their drawbacks and obtaining a measure that is simultaneously global and local. 
Furthermore, plugging the HIM kernel into a Support Vector Machine gives us a classification algorithm based on the HIM distance. 
First, we outline the theory underlying the metric construction. 
We introduce two diverse applications of the HIM distance and the HIM kernel to biological datasets. 
This versatility supports the adoption of the HIM family as a general tool for information extraction, quantifying difference among diverse instances of a complex system. 
An Open Source implementation of the HIM metrics is provided by the R package \textit{nettols} and in its web interface ReNette.
}

\section{Introduction}
\label{sec:intro}
\begin{figure}[!t]
\begin{center}
\includegraphics[width=1.0\textwidth]{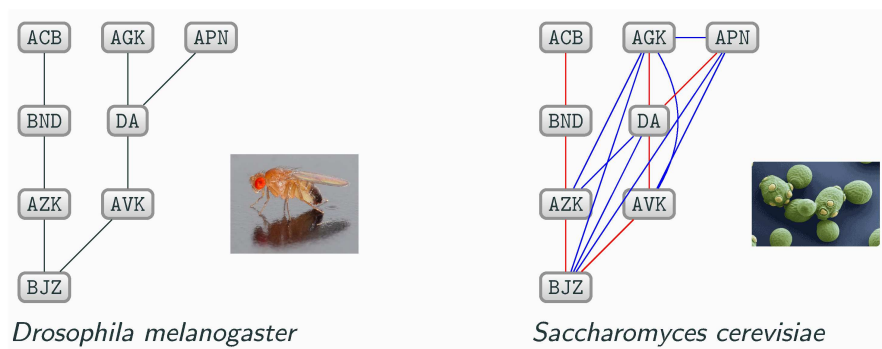}
\end{center}
\label{fig:dna}
\caption{A pair of similar subgraphs from a comparison of \textit{D. melanogaster} and \textit{S. cerevisiae} protein-–protein interaction network as shown in~\cite{cootes07identification}. Blue links are present only in the \textit{S. cerevisiae} subnet.}
\end{figure}

The paradigm shift towards complex systems science~\cite{barabasi12network}, stimulated by its recent theoretical and computational advance~\cite{csermely13structure,barabasi13network}, has paved the way for a parallel leap in computational biology by moving the focus from the differential gene expression analysis to differential network analysis (NetDA)~\cite{delafuente10from,ideker12differential}. 
Due to the heterogeneity in the NetDA process and potential ill-posedness of some of the involved functional operations~\cite{angulo15fundamental,baralla09inferring,meyer11verification}, a number of alternative approaches has appeared in the literature, with different strategies and aims~\cite{ha15dingo,gill10statistical,delafuente10from,sharan06modeling,ideker12differential,yoon12comparative,chuang07network,yang13network,pavlopoulos11using,barla12machine,barla13machine}.
For example, NetDA can be used to compare networks corresponding to different organisms, phenotypes or conditions. 
The subgraph of the protein-protein interaction network shown in Fig.~\ref{fig:dna} (from~\cite{cootes07identification}) is the same in terms of shared nodes for the fruitfly and the budding yeast. 
A group of links is shared by both instances of the subgraph, but the budding yeast network includes nine additional edges.
Clearly, when graphs to compare have a more complex structure, more sophisticated quantitative indicators are needed also to ensure a reproducibile analysis~\cite{ioannidis09repeatability}.
In general, the two key applications of NetDA are network comparison and network classification. 
Both can be framed in terms of similarity between graphs, which is best dealt by defining a distance. 
However, non-metric alternatives can be used~\cite{xiao08structure,dehmer13discrimination}, and even combinations of metric and statistical approaches~\cite{ruan15differential,koutra13deltacon}.

Here we propose to use the Hamming-Ipsen-Mikhailov (HIM) distance~\cite{jurman15him,jurman14him} first as the underlying metric for the NetDA framework, and also to induce a kernel for classification purposes.
The HIM metric linearly combines two distances, the Hamming~\cite{hamming50error,dougherty10validation,tun06metabolic,iwayama12characterizing,morris08specification} and the Ipsen-Mikhailov~\cite{ipsen02evolutionary}; the first is an edit distance, while the latter is a spectral measure.
These are the two most relevant families of graph distances: the edit distances are based on functions of insertion and deletion of matching links between the compared graphs, while the spectral measures are functions of the eigenvalues of one of the graph connectivity matrices.
The Ipsen-Mikhailov distance was chosen after a comparative review~ \cite{jurman11introduction}, while Hamming was selected as the simplest member of the edit family.
As a characterizing feature, HIM is a glocal distance that overcomes the drawbacks of local (edit) and global (spectral) metrics when separately considered. 
In fact, local functions disregards the overall network structure, while spectral measures cannot distinguish isospectral graphs.

NetDA based on the HIM distance has been used in metagenomics~\cite{zandona14metagenomic}, MEG neuroimaging~\cite{furlanello13sparse}, liver high-throughput oncogenomics~\cite{filosi14stability} and oncoimmunology~\cite{mina15tumor}. 
Moreover, the same method has found applicability also out of computational biology, \textit{e.g.}, socioeconomics~\cite{jurman15him} or even in multiplex network theory~\cite{jurman16metric}.
Here we present, after a brief summary of the main definitions, one application example in neurogenomics and one in developmental functional genomics. 
In the first example, we highlight and quantify weighted network dissimilarities among gene expression of brain tissues with different phenotypes (location, sex, health status), while in the latter we describe the trajectory of the binary developmental gene network in fruitfly across its different life stages.

Finally, we describe the CRAN R package \textit{nettools} and the web framework ReNette~\cite{filosi14renette}, which are available to implement NetDA projects.

\section{The HIM distance and kernel}
\label{sec:him}
\begin{figure}[!b]
\includegraphics[width=\textwidth]{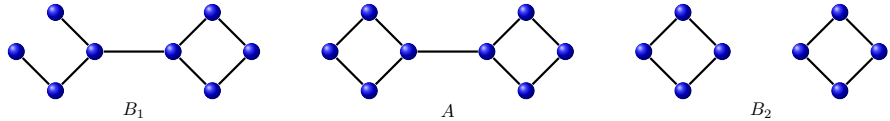}
\caption{Link equivalence for Hamming metric: $H(A,B_1)=H(A,B_2)$ although $B_1$ is connected while $B_2$ consists of two connected components.}
\label{fig:local}
\end{figure}

We recap hereafter the main definitions and results about the Hamming-Ipsen-Mikhailov (HIM) metric and kernel. 
The synthesis is based on the notations of Tab.~\ref{tab:notations}: for a fully detailed description, including mathematical proofs, we refer the reader to~\cite{jurman14him}.
\begin{table}[!t]
\caption{Notation and list of symbols}
\label{tab:notations}
\begin{center}
\begin{tabular}{|c|p{0.9\textwidth}|}
\hline
$\mathcal{N}_1,\mathcal{N}_2$ & Simple networks on $N$ nodes $\{z_i\}_{i=1}^n$\\
$A^{(1)}, A^{(2)}$ & Corresponding adjacency matrices, with $a^{(1)}_{ij}, a^{(2)}_{ij}\in\mathcal{F}$\\
$\mathcal{F}$ & Field $\mathbb{F}_2=\{0,1\}$ (unweighted case) or $[0,1]\subseteq\mathbb{R}$ (weighted case)\\
$\mathbb{I}_N$ & Identity matrix $\left( \begin{smallmatrix} 1&0&\cdots & 0 \\ 0&1&\cdots&0 \\ &\cdots \\ 0&0&\cdots &1  \end{smallmatrix} \right)$\\
$\textrm{1}_N$ & Unitary matrix $\left( \begin{smallmatrix} 1&1&\cdots & 1 \\ 1&1&\cdots&1 \\ &\cdots \\ 1&1&\cdots &1  \end{smallmatrix} \right)$ \\
$\textrm{0}_N$ & Zero matrix $\left( \begin{smallmatrix} 0&0&\cdots & 0 \\ 0&0&\cdots&0 \\ &\cdots \\ 0&0&\cdots &0  \end{smallmatrix} \right)$ \\
$\mathcal{E}_N$ & empty network (adjacency matrix $\textrm{0}_N$)\\
$\mathcal{F}_N$ & clique (adjacency matrix $\textrm{1}_N-\mathbb{I}_N$)\\
$\partial_g$  & degree of node $z_g$, $\partial_g=\partial(z_g)=\sum_{j=1}^N A_{gj}$\\ 
$D$ & Degree matrix $\left( \begin{smallmatrix} \partial_1&0&\cdots & 0 \\ 0&\partial_2&\cdots&0 \\ &\cdots \\ 0&0&\cdots &\partial_n  \end{smallmatrix} \right)$\\
$L$ & Laplacian matrix $D-A$, positive and semidefinite~\cite{chung97spectral}\\
$\textrm{spec}_L$ & Laplacian spectrum $\{0,\lambda_1,\lambda_2,\ldots,\lambda_N\}$, with $\lambda_1\leq\ldots\leq\lambda_N$ eigenvalues\\
$\omega_i$ & Vibrational frequencies $\sqrt{\lambda_i}$, solution of the ODE system $\displaystyle{\ddot{x}_i+\sum_{j=1}^N A_{ij}(x_i-x_j)=0}$ \cite{ipsen02evolutionary}\\
$\rho$ & Spectral density as sum of Lorentz distributions $\displaystyle{\rho(\omega,\gamma)=K\sum_{i=1}^{N-1} \frac{\gamma}{(\omega-\omega_i)^2+\gamma^2}}$\\
$K$ & normalization constant defined by $\displaystyle{\int_0^\infty \rho(\omega,\gamma)\textrm{d}\omega =1}$\\
$\gamma$ & half-width at half-maximum\\
$\overline{\gamma}$ & unique solution of $\displaystyle{\int_0^\infty \left[\rho_{\mathcal{E}_N}(\omega,\gamma)-\rho_{\mathcal{F}_N}(\omega,\gamma)\right]^2 \textrm{d}\omega=1}$ \cite{jurman14him}\\
\hline
\end{tabular}
\end{center}
\end{table}
The (normalized) Hamming distance~\cite{hamming50error,dougherty10validation,tun06metabolic,iwayama12characterizing,morris08specification} is the (local) simplest edit metric, counting the presence/absence of matching links:
\begin{displaymath}
\textrm{H}(\mathcal{N}_1,\mathcal{N}_2) = \frac{\textrm{Hamming}(\mathcal{N}_1,\mathcal{N}_2)}{\textrm{Hamming}(\mathcal{E}_N,\mathcal{F}_N)}  = \frac{1}{N(N-1)}\sum_{1\leq i\not = j\leq N} \vert A^{(1)}_{ij} - A^{(2)}_{ij} \vert\ .
\end{displaymath}
By definition, H ranges between 0 and 1, where
\begin{displaymath}
\textrm{H}=0\;\textrm{for $A^{(1)}=A^{(2)}$ and}\;
\textrm{H}=1\;\textrm{for $A^{(1)}+A^{(2)}=\textrm{1}_N-\mathbb{I}_N$} .
\end{displaymath}
Note that, for H, all links are equivalent regardless of their position within the network: for instance, in Fig.~\ref{fig:local}, both networks $B_1$ and $B_2$ differ from $A$ for just one link, and thus $H(A,B_1)=H(A,B_2)$, although $B_1$ is connected as $A$ while $B_2$ is not.
The Ipsen-Mikhailov distance~\cite{ipsen02evolutionary} is the (global) $L_2$ integrated difference of the Laplacian spectral densities:
\begin{displaymath}
\textrm{IM}(\mathcal{N}_1,\mathcal{N}_2)
= \sqrt{\int_0^\infty \left[\rho_{\mathcal{N}_1}(\omega,\overline{\gamma})-\rho_{\mathcal{N}_2}(\omega,\overline{\gamma})\right]^2 \textrm{d}\omega}\ .
\end{displaymath}
By definition, IM too ranges between 0 and 1, where
\begin{displaymath}
\textrm{IM}=0\;\textrm{for $\textrm{spec}(L^{(1)})=\textrm{spec}(L^{(2)})$ and}\;
\textrm{IM}=1\;\textrm{for $\{\mathcal{N}_1,\mathcal{N}_2\}=\{\mathcal{E}_N,\mathcal{F}_N\}$} .
\end{displaymath}
In fact, being a spectral measure, IM cannot distinguish isospectral (non isomorphic) networks.
\begin{svgraybox}
To overcome the drawbacks of both H and IM, we define their normalized cartesian product, the Hamming-Ipsen-Mikhailov distance: 
\begin{displaymath}
\textrm{HIM}_{\xi}(\mathcal{N}_1,\mathcal{N}_2) = \frac{1}{\sqrt{1+\xi}} \sqrt{ \textrm{H}^2(\mathcal{N}_1,\mathcal{N}_2) + \xi\cdot \textrm{IM}^2(\mathcal{N}_1,\mathcal{N}_2) },
\end{displaymath}
for $\xi\in [0,+\infty)$.
\end{svgraybox}
When $\xi$ is not close to the bounds $\{0, +\infty\}$ (and one of the factors becomes dominant), the impact of $\xi$ is minimal, and in general more relevant when $\textrm{HIM}_\xi$ is used as a kernel~\cite{furlanello13sparse}. 
Hereafter $\xi=1$ will be assumed, and the subscript $\xi$ omitted.
Again, HIM is bounded between 0 and 1, with
\begin{displaymath}
\textrm{HIM}=0\;\textrm{for $A^{(1)}=A^{(2)}$ and}\;
\textrm{HIM}=1\;\textrm{for $\{\mathcal{N}_1,\mathcal{N}_2\}=\{\mathcal{E}_N,\mathcal{F}_N\}$} .
\end{displaymath}
The HIM distance can be naturally extended to directed networks, by transforming it into an undirected bipartite graph through the procedure shown in~\cite{liu11controllability}.
\begin{svgraybox}
The HIM distance naturally induces a kernel via Gaussian (Radial Basis Function) map~\cite{cortes03positive,bolla13spectral}, to be used standalone or in a Multi-Kernel Learning framework to increase performance and enhance interpretability~\cite{kloft11lp}:
\begin{displaymath}
K(\mathcal{N}_1,\mathcal{N}_2) = e^{-\gamma\cdot\textrm{HIM}_\xi^2(\mathcal{N}_1,\mathcal{N}_2)}\ ,
\end{displaymath}
for a positive real number $\gamma$.
\end{svgraybox}
Although the HIM kernel is not positively defined in general for all $\gamma\in\mathbb{R}_0^+$, by results in~\cite{schoelkopf97support} it can be used in Support Vector Machines or other algorithms whenever $K$ is positive for the given training data, which is the case for all the examples shown in what follows.

\section{Application to -omic studies}
\label{sec:apps}

\subsection{The UKBEC dataset}
\label{sec:brain}
\begin{table}[!b]
\caption{Sample size of the UKBEC human brain dataset stratified by gender and tissue location (a) and by gender and age group (b). Region: the tissue location, Abbr.: abbreviation as in Fig.~\ref{fig:MF}, M: number of samples from male individuals, F: number of samples from female individuals. $a\sim b$ means $a< x\leq b$}
\label{tab:numerosity}
\begin{center}
\begin{tabular}{llr|r||llr|rclr|r||lr|r}
Region & Abbr. & M & F & Region & Abbr. & M & F & $\quad$ & Age & M & F & Age & M & F \\
\cline{1-8}\cline{10-15}
Cerebellar Cortex & CB &  95 &  35 & Frontal Cortex & FCX &  93 &  34 &&$< 32$      &86   & 39 & $58\sim 62$ & 117 & 20\\
Hippocampus & HC &  92  &  30 &  Medulla & Med& 88 &  31 && $32\sim 44$ & 130  &19  & $62\sim 68$ &72 & 29\\ 
Occipital Cortex &OCC&  94 &  35 &  Putamen & PUT& 96 & 33 && $44\sim 48$ & 74  & 24 & $68\sim 76$ &82 & 39\\ 
Substantia Nigra &SN&   73 &  28 &  Temporal Cortex & TCX& 86 &  33&&  $48\sim 53$ & 109  & 27 &$76\sim 83$ &66 & 56\\
Thalamus & Thal& 91 &  33 &  White Matter &WM&  97 &  34&& $53\sim 58$ & 101  & 20 & $\geq 83$ &68 & 53\\ 
\\
\multicolumn{8}{c}{(a)} & & \multicolumn{6}{c}{(b)}
\end{tabular}
\end{center}
\end{table}
The United Kingdom Brain Expression Consortium (UKBEC) hybridized on a Affymetrix Human Exon 1.0 ST Array (transcript version) 1213 human brain samples from 10 diverse regions. 
\begin{figure}[!t]
\includegraphics[width=1.0\textwidth]{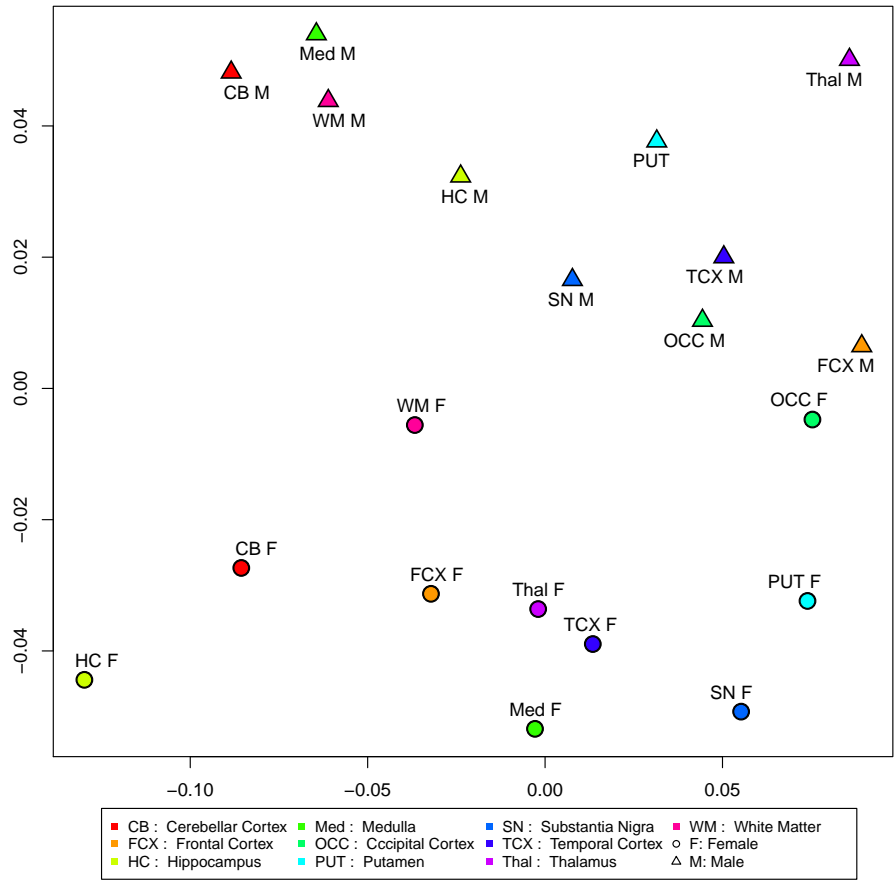}
\caption{Metric Multidimensional Scaling projection on two dimensions of all 190 mutual HIM distances between gene coexpression Brain Development networks stratified by gender and tissue locations.}
\label{fig:MF}
\end{figure}
Samples originated from 134 neurologically and neuropathologically normal individuals and were used in three studies aimed at better understanding gene expression differences~\cite{trabzuni14analysis,trabzuni13widespread,ramasamy14genetic}.
\begin{figure}[!t]
\begin{center}
\begin{tabular}{ccc}
\includegraphics[width=0.5\textwidth]{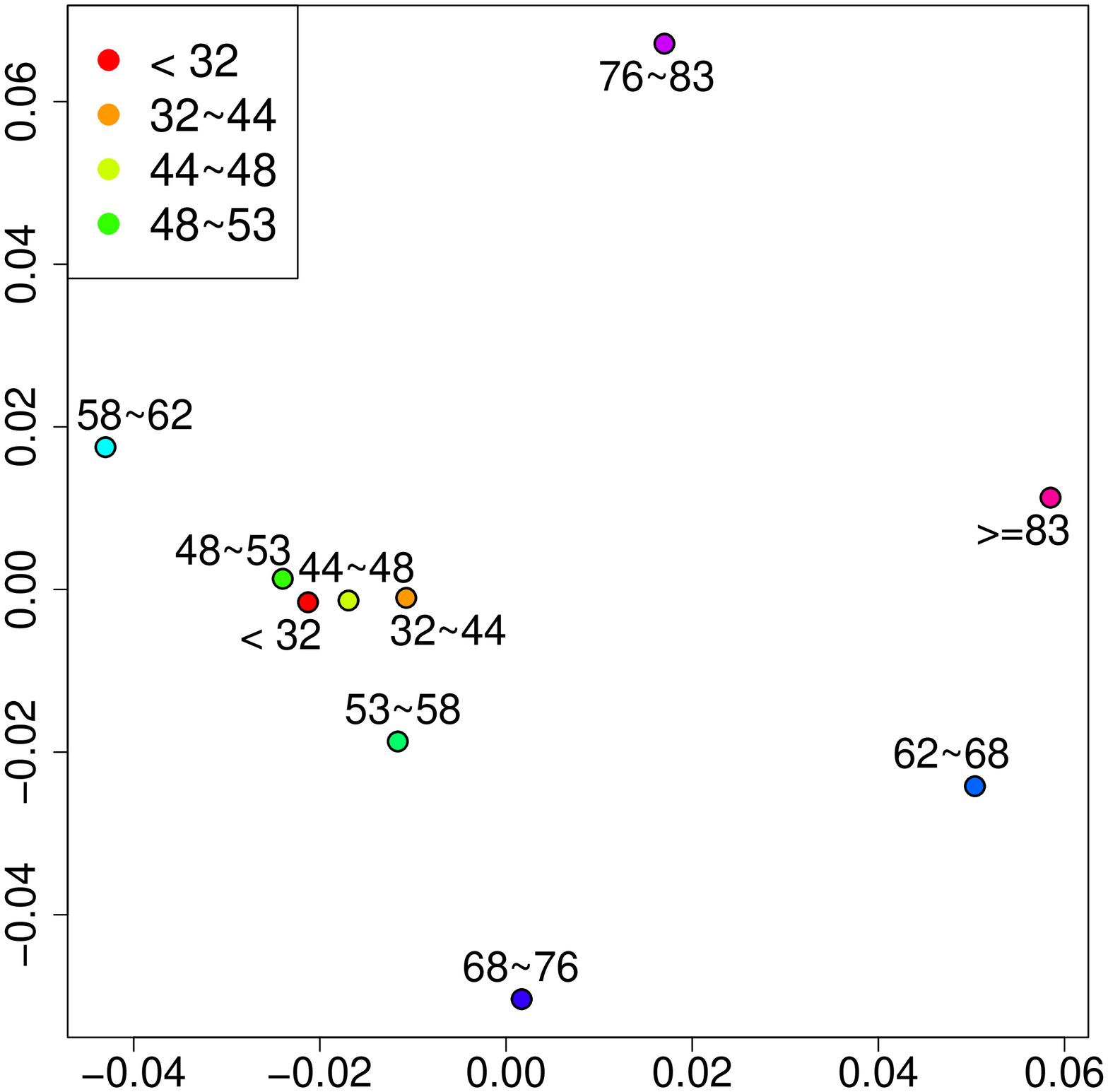} & \; &
\includegraphics[width=0.5\textwidth]{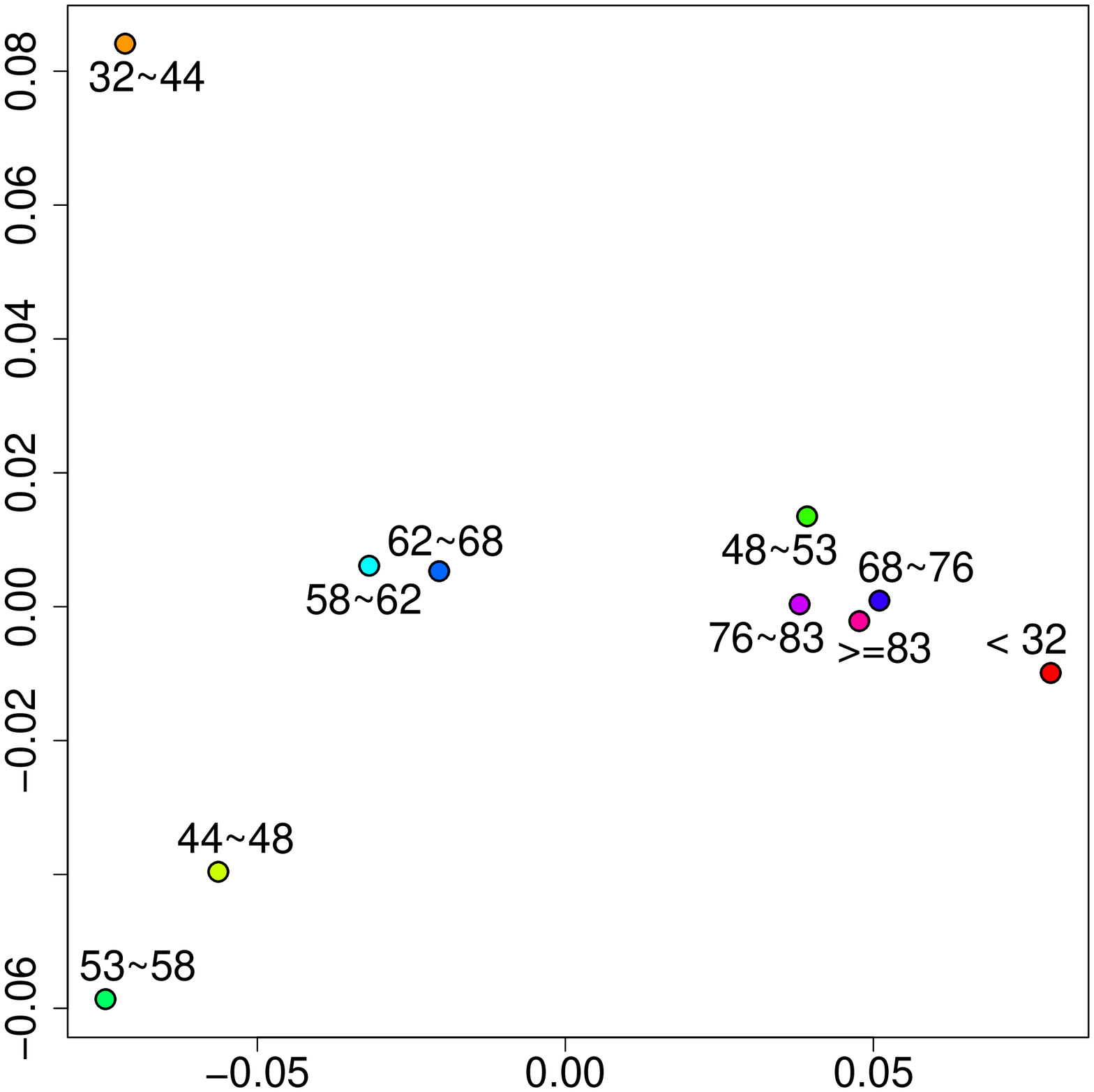} \\
(a) && (b)
\end{tabular}
\end{center}
\caption{Metric Multidimensional Scaling projection on two dimensions of all 45 mutual HIM distances between gene coexpression Brain Development networks stratified by age groups, separately for the male (a) and female (b) subjects. $a\sim b$ means $a< x\leq b$.}
\label{fig:aging}
\end{figure}
Data details about sample stratification according to sex and tissue location are listed in Tab.~\ref{tab:numerosity}(a).
Here, this dataset\footnote{available as GEO46706 at \url{http://www.ncbi.nlm.nih.gov/geo/query/acc.cgi?acc=GSE46706}} is used to build the absolute Pearson coexpression networks corresponding to different phenotypes defined on the 50 genes involved in the BRAIN\_DEVELOPMENT (GO:0007420, GSEA M7203) pathway\footnote{available at \url{http://software.broadinstitute.org/gsea/msigdb/cards/BRAIN_DEVELOPMENT}}, corresponding to 1012 probes on the Affymetrix Human Exon 1.0 ST Array platform\footnote{the platform has no probes for the 51st gene of the pathway, VCX3A}.

First, we consider planar projections of all the mutual HIM distances between networks with shared nodes based on the metric multidimensional scaling (mMDS)~\cite{mardia78some,cox01multidimensional} in Fig.~\ref{fig:MF}.
The mMDS plot shows the mutual HIM distances with networks stratified for both sex and tissue location. 
Citing the authors, the study in~\cite{trabzuni13widespread} ''provides unequivocal evidence that sex-biased gene expression in the adult human brain is widespread in terms of both the number of genes and range of brain regions involved''. 
In our analysis, the result is numerically confimed by the major effect emerging at the gene coexpression level (Fig.~\ref{fig:MF}): male and female networks can be linearly separated in the mMDS space, with large HIM distances between both inter- and intragender tissue locations.
In particular, intragender HIM distances among different tissue regions are larger for the female samples (range [0.112,0.232], median 0.146) than for the male (range [0.077,0.200], median 0.118), with statistical significance (t-test p-value $1.9\cdot 10^{-4}$).

In Fig.~\ref{fig:aging}, we show instead the mMDS projections for the mutual HIM distances of the coexpression networks built separately for male and female subjects, partitioned in 10 age groups: the sample size for each network is listed in Tab.~\ref{tab:numerosity}(b).
While the plot for the females does not show any global pattern, for males the first 5 groups (age $<58$ y) have small mutual HIM distances and they result clustered together. 
On the other hand, the five older male groups are both mutually distant and distant from the younger subjects cluster, too.
In this dataset, the small sample size in the female subgroup may be a relevant source of noise for some of the age groups, \textit{e.g.} the $32\sim 44$.
Our results are consistent with findings obtained with different data and methodology by Berchtold and colleagues in~\cite{berchtold08gene}, suggesting the existence of a global pattern of gene expression change associated with brain aging, more evident from the sixth decade onward, with different evolutions between males and female, with larger variations in male subjects.  
Biologically, this is due to a wider global decrease in males in the catabolic and anabolic capacity with aging, mainly in genes linked to energy production and protein synthesis and transport~\cite{berchtold08gene}.

\subsection{The \emph{D. Melanogaster} development dataset}
\label{sec:fruitfly}
\begin{figure}[!ht]
\begin{center}
\begin{tabular}{cc}
\begin{tabular}{c}
t=1\\
\includegraphics[width=0.3\textwidth]{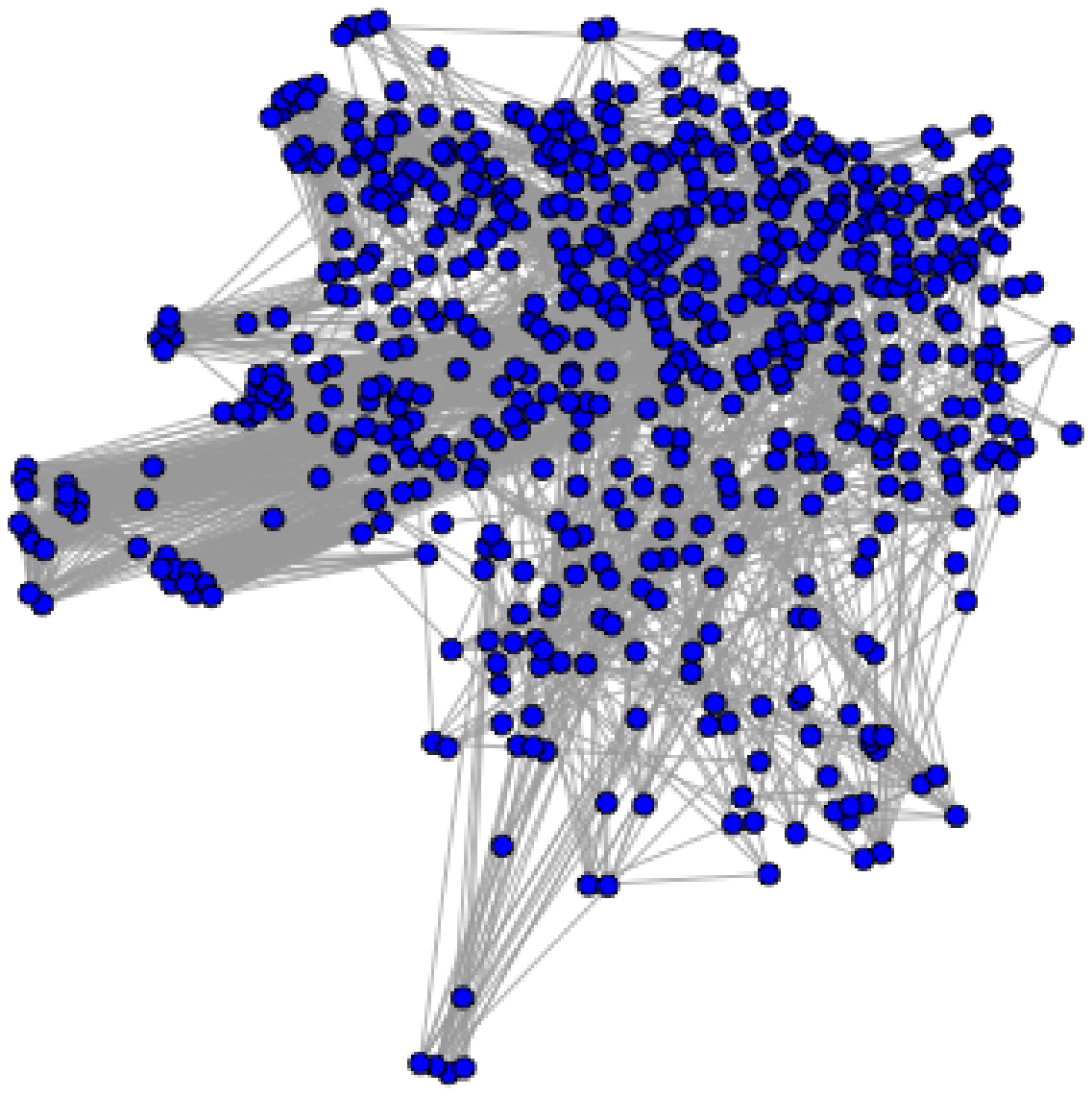} \\
t=20 \\
\includegraphics[width=0.3\textwidth]{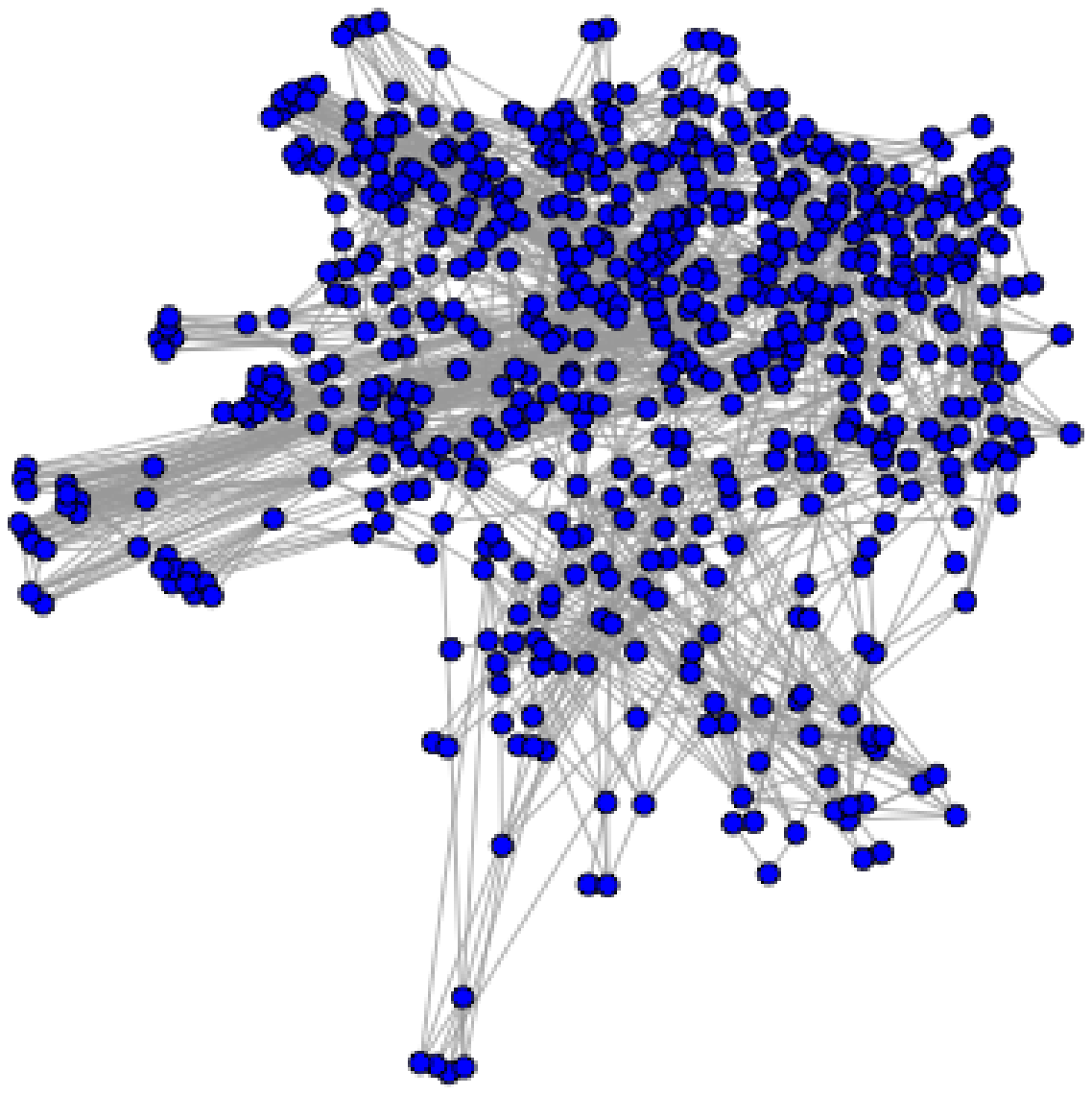} \\
t=35 \\
\includegraphics[width=0.3\textwidth]{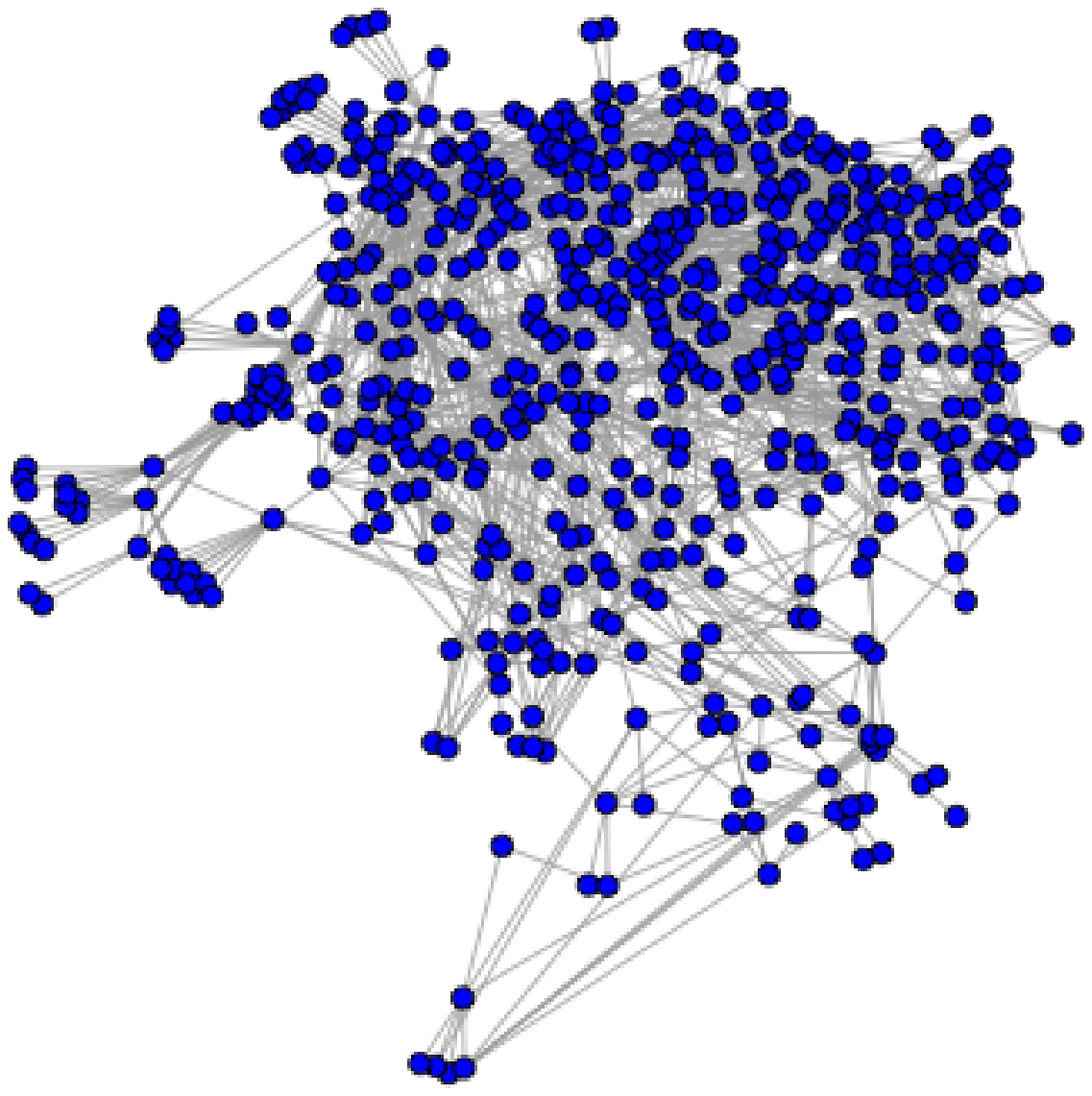} \\
t=66 \\
\includegraphics[width=0.3\textwidth]{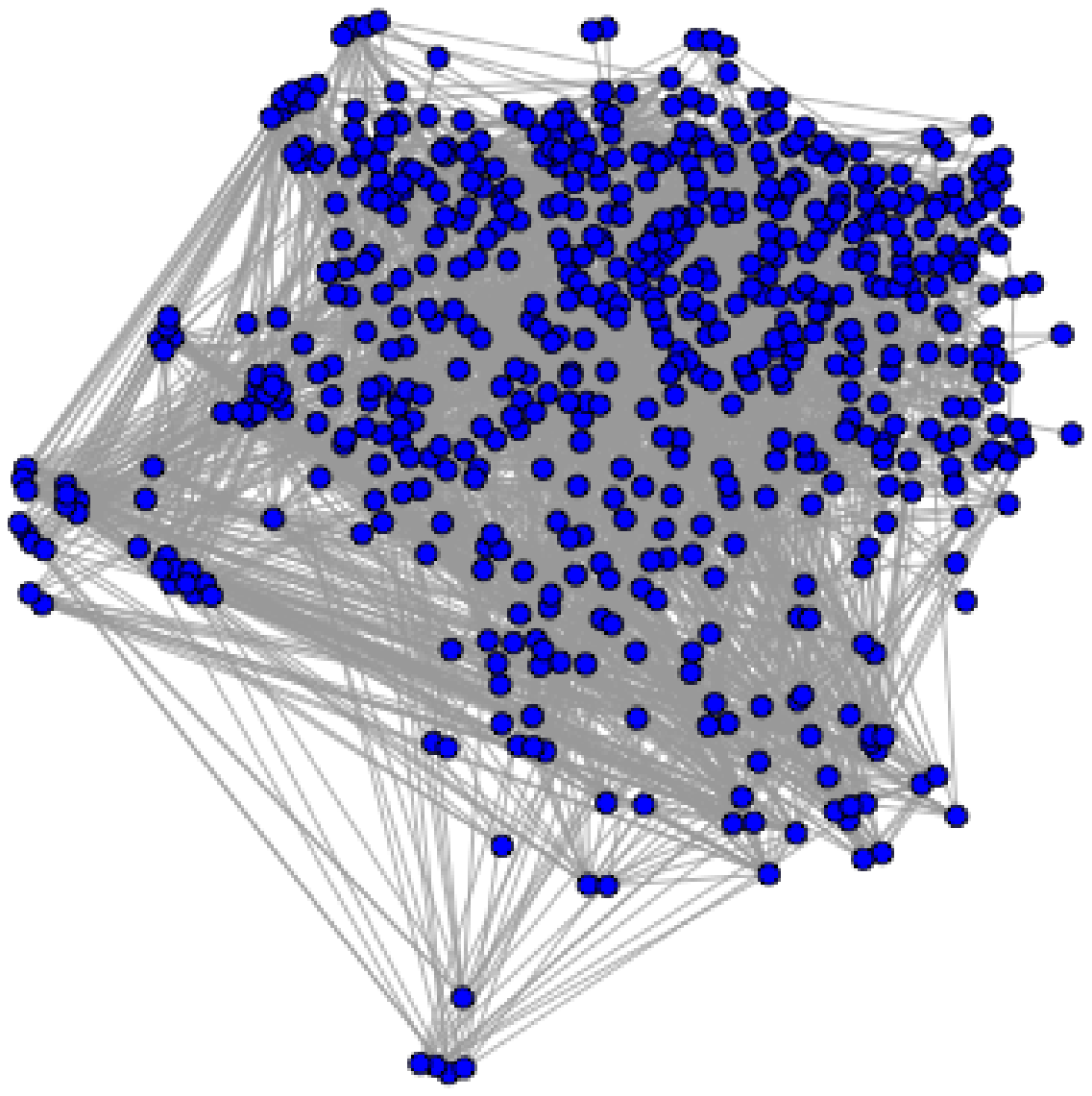} \\
\\
(a)
\end{tabular}
&
\begin{tabular}{r}
\includegraphics[width=0.6\textwidth,angle=-90]{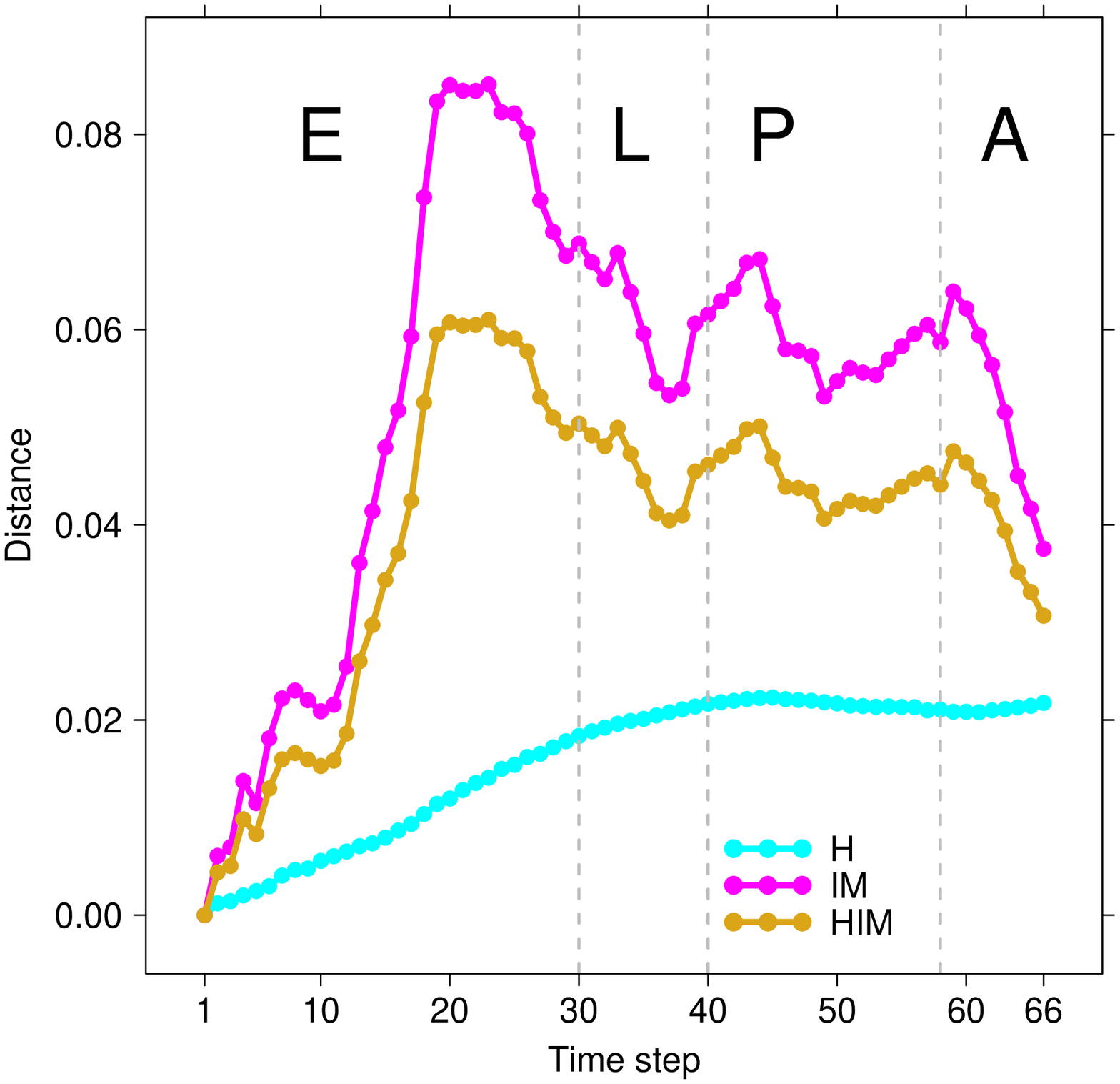}\\
\multicolumn{1}{c}{(b)}\\
\\
\\
\\
\includegraphics[width=0.55\textwidth]{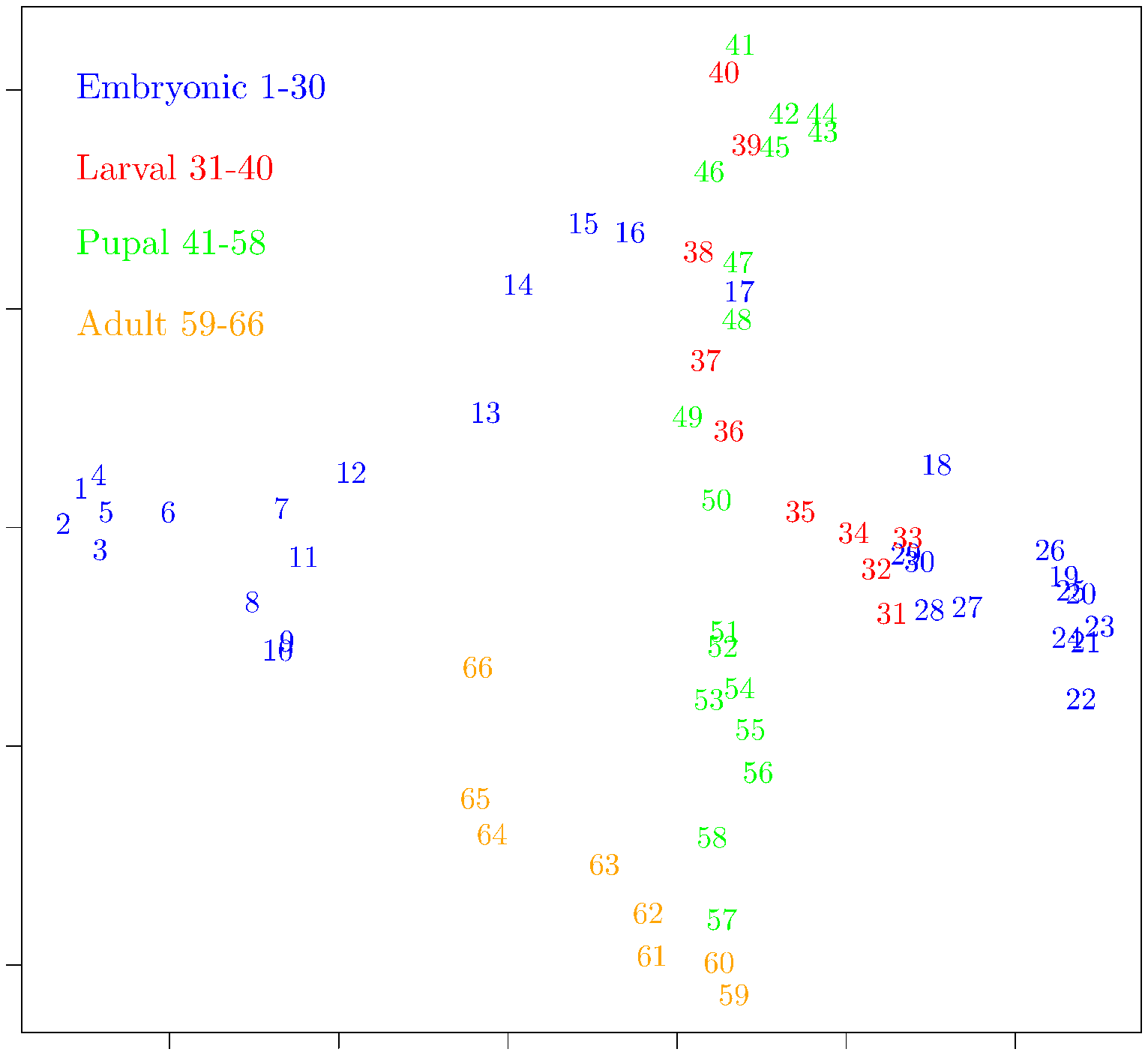}\\
\\
\multicolumn{1}{c}{(c)}
\end{tabular}
\end{tabular}
\caption{\textit{D. melanogaster} development network dataset. (a) Keller interaction network $N_i$ for the \textit{D. melanogaster} development genes at the time points $i=1,20,35,66$. (b) Evolution of H (cyan), IM (magenta) and HIM (goldenrod) distances network time series across 66 time points in the 4 stages Embryonic (E), Larval (L), Pupal (P) and Adult (A). (c) Metric multidimensional scaling planar projection of the mutual HIM distances between the 66 networks $N_i$, colored according to the developmental stage Embryonic (blue), Larval (red), Pupal (green) and Adult (orange).}
\label{fig:droso}
\end{center}
\end{figure}

In~\cite{kolar10estimating}, Kolar and colleagues applied the Keller algorithm to infer the gene regulatory networks of \textit{Drosophila melanogaster} from a time series of gene expression data measured during its full life cycle, originally published in \cite{arbeitman02gene}.
They followed the dynamics of 588 development genes along 66 time points spanning through four different stages (Embryonic -- time points 1-30, Larval -- t.p. 31-40, Pupal -- t.p. 41-58, Adult -- t.p. 59-66), constructing a time series of inferred networks $N_i$\footnote{publicly available at \url{http://cogito-b.ml.cmu.edu/keller/downloads.html}}: in Fig.~\ref{fig:droso}(a) we show four instances of the $N_i$ networks, at different timing.

As a first step in the quantitative NetDA of this dataset, we measure the HIM distance between each $N_i$ and the initial network $N_1$: the resulting distance time series is shown in Fig.~\ref{fig:droso}(b).
The largest variations, both between consecutive terms and with respect to the initial network $N_1$, occur in the embrional stage (E). 
In particular, the HIM distance grows until time point 23; next networks get closer again to $N_1$, showing that the interactions of the selected 588 genes in the adult stage are more similar to the corresponding net of interaction in the embrional stage, rather than in the other two stages.
Moreover, while the Hamming component ranges between $0$ and $0.0223$, the Ipsen-Mikhailov distance has $0.0851$ as its maximum, indicating an higher variability of the networks in terms of structure rather than matching links.

Then we computed all 2145 HIM distances $\textrm{HIM}(N_i,N_j)$, and we projected them on a 2D mMDS representation, shown in Fig.~\ref{fig:droso}(c).
Interestingly, the networks for the Embryonal stage split into two clusters (before and after time points 17), and the Embryonal and Pupal stages are orthogonal in this representation.

Moreover, the Adult stage networks form a cluster well separated from the other nets, with the Larval stage graphs mixing with the Pupal and late Embryonal stages.
Finally, a Support Vector Machine classifier with HIM kernel was developed with the \textit{kernlab} package in R, with a 5-fold Cross Validation with $\gamma=10^3$ and $C=1$.
The classifier reached accuracy 0.97 in discriminating Embryonic and Adult networks from Larval and Pupal. 
Similarly, in the same setup, perfect separation is reached between Embryonic and Adult stages for all values of $\gamma > 10^3$.

\FloatBarrier
\section{Conclusion}
\label{sec:conclusion}
The interest of the HIM metric is its global/local approach: by combining edit and spectral distance types, we overcome the drawbacks of the two distance components.
The two applications in functional hig-throughput -omics presented support the effectiveness of the approach.
The strategy of a NetDA based on the HIM distance offers a reproducible method: the metric gives a completely quantitative assessment of the differences among networks (on shared nodes) as well as a scalar product for kernel learning machines. 

Operatively, we provide an Open Source implementation of the HIM distance with the R package \textit{nettools} available on CRAN and GitHub\footnote{\url{https://github.com/MPBA/nettools.git}}, and in the web interface ReNette~\cite{filosi14renette}\footnote{\url{http://renette.fbk.eu}}. 
In particular, ReNette includes a complete pipeline for NetDA, integrating a comprehensive collection of tools for network inference, network comparison and network stability analysis~\cite{filosi14stability} through queue-based submission system and  asynchronous task management.
The software is already configured for usage on multicore workstations, on high performance computing (HPC) clusters and on the cloud, to deal with the extraction of the Laplacian spectrum, which represents the computational bottleneck of the algorithm.
\bibliographystyle{spbasic}
\bibliography{jurman16differential}
\end{document}